

\magnification=1200
\baselineskip=16pt
\hskip 10cm
ICN-UNAM 9415
\vskip2pc
\centerline{\bf THE CONSTRAINTS IN SPHERICALLY SYMMETRIC}
\centerline{\bf  CLASSICAL GENERAL RELATIVITY I}
\vskip2pc
\centerline{\bf OPTICAL SCALARS, FOLIATIONS, }
\centerline{\bf BOUNDS ON THE CONFIGURATION SPACE VARIABLES}
\centerline{\bf AND}
\centerline{\bf THE POSITIVITY OF THE QUASI-LOCAL MASS}
\vskip2pc
\centerline {\bf Jemal Guven$^{(1)}$ and Niall \'O Murchadha$^{(2)}$ }
\vskip1pc
\it
\centerline {$^{(1)}$Instituto de Ciencias Nucleares}
\centerline {Universidad Nacional Aut\'onoma de M\'exico}
\centerline {A. Postal 70-543. 04510 M\'exico, D. F., MEXICO} \rm
\centerline{(guven@roxanne.nuclecu.unam.mx)}
\vskip2pc
\it
\centerline {$^{(2)}$ Physics Department}
\centerline {University College Cork}
\centerline {Cork, IRELAND} \rm
\centerline{(niall@iruccvax.ucc.ie)}
\vfill\eject
\centerline{\bf Abstract}
\vskip1pc
{\leftskip=1.5cm\rightskip=1.5cm\smallskip\noindent
This is the first of a series of papers in which we
examine the constraints of spherically symmetric
general relativity with one asymptotically flat region.
Our approach is manifestly invariant under
spatial diffeomorphisms, exploiting both traditional
metric variables as well as the optical scalar variables
introduced recently in this context.
With respect to the latter variables, there exist
two linear combinations of the Hamiltonian and momentum constraints
one of which is obtained from the other by time reversal.
Boundary conditions on the  spherically symmetric
three-geometries and extrinsic
curvature tensors are discussed. We introduce a one-parameter
family of foliations of spacetime involving
a linear combination of the two scalars characterizing a
spherically symmetric extrinsic curvature tensor. We can
exploit this gauge to express one of these scalars in terms
of the other and thereby solve the radial momentum constraint
uniquely in terms of the radial current.
The values of the parameter yielding potentially globally
regular gauges correspond to the vanishing of a timelike vector in the
superspace of spherically symmetric geometries.
We define a quasi-local mass (QLM) on spheres
of fixed proper radius which provides observables of the theory.
When the constraints are satisfied the QLM can be expressed
as a volume integral over the sources and is positive. We provide two
proofs of the positivity of the QLM. If
the dominant energy condition (DEC) and the constraints are
satisfied positivity can be established in a manifestly
gauge invariant way. This is most easily achieved exploiting the
optical scalars. In the second proof we specify the foliation.
The payoff is that the weak energy condition replaces
the DEC and the Hamiltonian constraint replaces the
full constraints. Underpinning this proof is
a bound on the derivative of the
circumferential radius of the geometry  with respect to its
proper radius. We show that, when the DEC is satisfied, analogous
bounds exist on the  optical scalar variables and, following on from
this, on the extrinsic  curvature tensor.
We compare the difference between the
values of the QLM and the corresponding material energy to prove that
a reasonable definition of the gravitational binding energy is
always negative. Finally, we summarize our
understanding of the constraints in a tentative
characterization of the configuration space of the
theory in terms of closed bounded trajectories
on the parameter space of the optical scalars.\smallskip}

\vfill
\eject

\noindent{\bf I INTRODUCTION}
\vskip1pc

To identify the independent dynamical degrees of freedom of the gravitational
field in general relativity it is useful to cast the theory in Hamiltonian
form [1]. This means that the gravitational field must be viewed, not as a
fixed four-dimensional object, but rather as a sequence in `time' of
Riemannian three-geometries. Thus we think of specifying some initial
configuration of sources and gravitational field and letting it evolve.
As is well known, this initial data cannot be specified arbitrarily,
it must satisfy the Einstein constraint equations.
These constraints only contain the source energy density and momentum density.
They do not depend on the equations of state. Of course, if
we wanted to track the evolution of the system we
would need to provide a more detailed specification of the sources,
including these equations of state.

In this paper we will focus on the solution of the
classical constraints and the identification of those features
of the theory which depend only on the initial data in the
simplified setting of spherically symmetric general relativity
with our sight set on the quantum theory.

A remarkable consequence of the diffeomorphism invariance of
general relativity is that, in a sense,
the constraint equations are all there is to the theory.
For if the constraints are satisfied at
all times and the sources are completely specified, then the
evolution equations follow [2]. For this reason, the solution of
the constraints should be viewed as
much more than a prerequisite to the solution of the dynamical problem.
Once a point in the classical configuration space ({\it i.e.} a
solution of the constraints) has been identified, its subsequent
evolution is implicitly defined. The structure of this
space is, of course, highly non-trivial.

The most developed classical approach to the
solution of the constraints has been the
conformal geometry approach  pioneered by York and co-workers
in the seventies. This was very successful in settling formal questions
such as the existence and uniqueness of solutions [3].
However, beyond this formal level, it is extremely difficult
to piece together the structure of the
configuration space of the full theory outside the
domain of perturbation theory.
Unfortunately, this is the framework on which the canonical
quantization of the theory is based. Thus, until this is done
any claims we make about the quantum theory
must necessarily be taken with a grain of salt.

One regime in which the problem simplifies, without
sacrificing all local dynamical degrees of freedom (such as we do
in homogeneous relativistic cosmologies), is
when the geometry as well as the material sources are
spherically symmetric [4,5,6,7]. In such a system, all the true local
dynamical degrees of freedom reside in the sources. There are
no independent local gravitational dynamical degrees of freedom.
The sources, however, generate a `gravitational potential',
a kinematical object, which in turn interacts on them.
The dynamics of matter associated with this
potential can be extremely non-trivial, a point convincingly
demonstrated by the recent controversy created by Choptuik's
numerical simulations of the collapse of a massless
scalar field [8].

There are only four topologies compatible with spherically symmetric
initial data that is defined on a three-manifold.
The manifold can be $R^3$, with a regular center and one end, just like
ordinary flat space; it can be $S^2 \times R^1$, with two ends and no
center as with the spatial slice through extended Schwarzschild
spacetime;  it can be $S^2 \times S^1$ which is the spherically
symmetric  torus or it can be $S^3$, the three-sphere [9].
We limit the discussion to the first case, {\it i.e.},
to geometries possessing one asymptotically flat region deferring the
examination of spherically symmetric inhomogeneous
cosmologies and the double ended case to future publications.

The boundary conditions associated with the given topology
play an important role. In the case we will study,
the  only boundary condition we need to implement is the
regularity (or the degree of singularity) at the base of the
spatial geometry. Technically this is because
the Hamiltonian constraint is a singular ordinary differential
equation at this  point. On one hand, this imposes an extraordinary rigidity
on the solution, making it unique. On the other it provides the
mechanism, when the energy density is appropriately large, which
allows  singularities to occur in the geometry. If the material sources
are suitably localized (as we will always assume)
the constraints will automatically
steer the geometry to asymptotic flatness if no singularity
intervenes. In closed cosmologies, the non-singular closure
of the spatial geometry imposes integrability
conditions on the sources it contains.

The initial data for the gravitational field consists of two parts,
the intrinsic geometry of the three-manifold and an
extrinsic curvature tensor which describes how this
three-manifold is embedded into a four dimensional spacetime.
The solution of the constraints involves the
implementation of gauge conditions. One of these conditions
involves the specification of how these three-manifolds foliate spacetime.
There are two ways of doing this within the canonical context;
intrinsically, where
the foliation is determined by placing some restriction on the
three-metric, for example, that it be flat; and extrinsically, where
some condition is placed on the extrinsic curvature, for example,
that its trace vanishes (the maximal slicing condition). For
any choice one must show that it is compatible
with the constraints and that it can be used as an evolution
condition. It has been found that doing this
extrinsically is invariably better than doing it intrinsically.
In view of this we will only consider
extrinsic slicing conditions.

The remaining gauge condition concerns the specification of the
spatial coordinate system. The point of view we will adopt in this paper
is that it is not necessary, at least at the level of the constraints,
to make an explicit spatial coordinate choice.
The justification for this is the fact that
there are two invariant linear measures of the spherically symmetric
geometry, the circumferential radius $R$, and the proper
radial length $\ell$, and the constraints come ready
cast in terms of derivatives of $R$ with respect to $\ell$.

It is natural that the gauge which fixes the foliation is the gauge which
should be tackled first. Fix the foliation, then fix
coordinates on the hypersurfaces picked out by this foliation.
Having said this, it is only fair to also point out
that the choice of gauge which simplifies the
solution of the constraints most dramatically is implemented
most efficiently by  inverting this order,  exploiting
the circumferential radius as the radial coordinate
and then foliating spacetime by the so called polar gauge\footnote *
{This is the gauge exploited in refs. [4,5] and [6].}
In this gauge, not only does the extrinsic curvature
quadratic miraculously fall out of the Hamiltonian constraint so that
it mimics its form at a moment of time symmetry, but the constraint
is then also exactly solvable. Furthermore, the momentum  constraint
reduces to an algebraic equation which permits the
non-vanishing extrinsic curvature component to be
determined locally in terms of the material
current.  What is unfortunate is that both the foliation and the
spatial coordinate system break down catastrophically
when the geometry possesses an apparent horizon.
This corresponds to the vanishing of
one or the other of $\Theta_\pm$,
the divergence of the future and past pointing
outward directed null rays on a metric two-sphere
at fixed proper radius [10].

For the purpose of examining observable effects in the
classical theory it is sufficient to truncate
the geometry at the horizon if it possesses one, and
place appropriate boundary conditions there. Even if the
formation of the horizon is a consequence of
physical processes occuring in its interior once formed the
details of the interior physics can have no observable consequences in the
exterior. In the quantum theory, however,
we know that we are not always
at liberty to truncate the theory in this way.

On one hand, a process like the Hawking effect can be
understood in terms of the polarization of the vacuum in the exterior
neighborhood of the event horizon [11]. In the approximation
in which the back-reaction on the geometry can be ignored
the techniques of quantum field theory on a given curved
background spacetime apply. What is beyond the scope of any
approximation which truncates the geometry at the horizon is the prediction
of the final state of the black hole.

There are other processes, however, still below the
Planck regime, such as tunnelling from
a configuration without an apparent horizon to a configuration with
an apparent horizon, in which the existence of classically
inaccessible regions of the spatial geometry can have dramatic
consequences in the quantum theory [12,13].

The above processes in quantum gravity are both
semi-classical in nature. In the Planck regime, however,
we do not even possess an unambiguous classical lump to start with.
Furthermore, the very definition of an apparent horizon involves both
the intrinsic and the extrinsic geometry (or equivalently the
momentum conjugate to the intrinsic geometry) which are
not simultaneously observable in the quantum theory.

If we adhere to a configuration space
consisting of metric variables, for the canonical
quantization  of the model we will need to catalog all
possible solutions satisfying the constraints with or without
apparent horizons. The only way to mend the situation to accomodate
the polar gauge would be to introduce the gauge patch by patch
between successive horizons. We do not examine this possibility
here because it would be almost impossible to implement
in the quantum theory.

A canonical change of variables, from the traditional
metric phase space variables, to the optical tensor variables
defined on a foliating sequence of closed two-dimensional
hypersurfaces embedded in the three-geometry provides an extremely
useful alternative description of the initial data when
the geometry is spherically symmetric. In this case, when the
two dimensional hypersurfaces are also spherically symmetric, the
optical tensors reduce to the two scalar
quantities, $\Theta_\pm$. The vanishing of $\Theta_+$  corresponds to
a future apparent horizon, and the vanishing of $\Theta_-$ to a
past apparent horizon. Thus, by adopting this variable to characterize
the  configuration space, we sidestep the difficulty  inherent in the
metric variable description of apparent horizons in the quantum theory.
These variables are a linear combination of intrinsic and
extrinsic quantities [14].
Most importantly, is that, when cast with respect to
the optical scalars, we can replace the Hamiltonian constraint and the
momentum constraint by a pair of quasi-linear first-order equations,
one of which is the time reversal of the other  [15] and which are
entirely equivalent to the original constraints.

In Sect.3 we return to the metric variables in a search
for a globally valid foliation. We introduce a one-parameter family of
foliations corresponding to the vanishing of some linear
combination of the two independent scalars characterizing the
extrinsic curvature in a spherically symmetric geometry.
Each such gauge corresponds to a ray in superspace [16].
The physically acceptable foliations correspond to timelike directions.
Maximal slicing is one of these. With the optical variables, this is
the natural choice of gauge. However, we find that there
are other unexpected parameter values possessing attractive features.
One of the lightlike directions in superspace bounding the valid gauges
corresponds to polar gauge. In the gauge defined by the
other lightlike direction, the Hamiltonian constraint also
mimicks its form at a moment of time symmetry. As such, it
is worth considering more closely.
Minimal surfaces in this gauge, however, do not coincide with apparent
horizons. What is more serious, the foliation is
not suitably aymptotically flat.

One of the most remarkable results in general relativity is the positivity of
the
Arnowitt-Deser-Misner (ADM) mass, the result of a conspiracy occurring
at the level of the constraints which ensures that the
Hamiltonian of the theory is positive definite.
When the spacetime geometry is spherically symmetric
there also exists a quasi-local  mass (QLM) which is
positive and reduces to the ADM mass at
infinity in an asymptotically flat geometry [17].
Attempts to find an analogous quantity
when this symmetry is relaxed which is also positive have failed.

In Sect.4, the QLM of a spherically symmetric geometry is
introduced as an integral over a spherical surface
of fixed proper radius of a spacetime scalar quantity,
and, as such, an observable of the theory by any reasonable criterion [18].
When an appropriate linear combination of the constraints
is satisfied the QLM can be expressed as a volume integral
over the sources. An equivalent expression was derived by
Fischler et al. in ref.[17] (see also [19]). The QLM
thereby provides a very useful first integral of the
constraints.

In Sect.5 we provide two proofs of the positivity of the QLM when the
geometry is regular.

If the dominant energy condition (DEC) ([10]) and the constraints are
satisfied positivity can be established in a manifestly
gauge invariant way [15]. This is achieved remarkably
easily by exploiting the
optical scalars. We comment on the approach to a singularity when the QLM
is negative. The second proof is weakly gauge dependent. However, it has the
peculiar property of permitting us to
replace the DEC by the weak energy condition and
ignore the momentum constraint when we use a linear extrinsic curvature
foliation of spacetime.

Both of these positivity proofs arise as simple corollaries
to the existence of appropriate bounds on the phase space variables;
in the former case an upper bound on the product of the
optical scalars [15]; in the latter, by the bound on the derivative
of the circumferential radius of the geometry  with respect to its
proper radius: $-1 < \partial_\ell R \le 1$.
This bound has a simple geometrical interpretation
in terms of the embedding of the geometry in flat $R^4$.

When the DEC is satisfied and the geometry is regular,
additional bounds can be placed on the
values assumed by the optical scalars,
which, in turn, imply a bound on the
extrinsic curvature. Considering the
identification of these variables as the momenta
conjugate to the spatial metric this is a particularly intriguing
result. These bounds assume a particularly simple form when the
foliation is maximal. It is not clear what role these bounds
play in the theory. They do not appear to be related
directly to the positivity of the QLM. It is possible, however,
that they will prove to be more fundamental.
The derivations we provide in Sect.6 are more
economical and the bounds tighter than those derived in [15].

In Sect 7. we compare the values of the QLM and the
material energy. Any reasonable defined measure of the
gravitational binding energy should always be negative. In particular,
we demonstrate that the naive definition consisting
of the difference between the QLM and the material energy
is negative when the foliation of spacetime is maximal.

In Sect.8 we summarize our understanding of the
constraints in terms of the optical scalars. This
is done by associating with each regular solution of the constraints
a closed bounded trajectory on the parameter space of
the optical scalars. The set of all such trajectories
can be identified as the phase space of the theory.

Because of the importance of instantons
in the semi-classical approximation, we will also occasionally
comment on the form of the constraints in Euclidean signature relativity.

We finish with a summary and an outline of subsequent papers [20][21].
\vskip2pc
\noindent{\bf 2 THE CONSTRAINTS}
\vskip1pc

\noindent{2.1 The Constraints in terms of metric variables}
\vskip1pc
Initial data for the gravitational field
in general relativity consist of a spatial metric
$g_{ab}$ and an extrinsic curvature tensor $K_{ab}$ which satisfy
the constraints [1,2]:

$$K^2-K^{ab}K_{ab} + {\cal R} = 16\pi \rho\,,\eqno(2.1a)$$
and

$$\nabla^bK_{ab}-\nabla_a K = - 8\pi J_a\,.\eqno(2.1b)$$
${\cal R}$ is the three-scalar curvature constructed with $g_{ab}$.
The three-scalar $\rho$ is the material energy per unit physical three-volume.
The three-vector $J^a$ is the corresponding current.
When the signature of the
spacetime metric is made positive definite the sign of the quadratic
terms in the  extrinsic curvature appearing in Eq.(2.1a) is reversed.

We will examine spherically symmetric spacetime geometries.
The only non-trivial space-time directions are
the radial and time directions orthogonal to the
orbits of rotations and the geometry can be
described by a line element of the form

$$ds^2= -(N^2(r,t)-\beta^2(r,t)) dt^2 + 2 \beta(r,t) dt dr +
{\cal L}(r,t)^2dr^2+R^2(r,t)d\Omega^2\,.\eqno(2.2)$$
The spatial geometry at constant $t$,
we parametrize by two functions ${\cal L}$ and $R$
of the radial coordinate $r$. $N$ and $\beta$ are
respectively the lapse and the radial shift.
The scalar curvature of the spatial geometry is now given by

$${\cal R}=-{2\over R^2}
\Bigg[2 \Big(R R^\prime \Big)^\prime -R^{\prime 2}
-1 \Bigg]\,.\eqno(2.3)$$
We introduce the notation $'$ to denote
the derivative with respect to the proper radius $\ell$  defined by
$d\ell ={\cal L}dr$. When radial derivatives are taken with
respect to $\ell$, ${\cal L}$ no longer appears
explicitly in the constraints. The requirement that this
condition be preserved under the dynamical evolution of the
spatial geometry will determine $\beta$ implicitly.
In general $\beta$ will {\it not} be zero.

The other invariant geometrical measure of a spherically
symmetric geometry is the circumferential radius $R$.
The identification of $r$ with $R$ is the radial coordinate
choice which is most
frequently adopted.\footnote * {Indeed, with respect to this
choice of radial coordinate and a foliation by $K_R=0$,
the Hamiltonian constraint reduces to an
exactly solvable linear first order differential equation for ${\cal L}^{-1}$.
In addition, $\beta=0$.}
The difficulty, however, is that this identification breaks down
wherever $R^\prime=0$ which is the condition that the
two-surface of constant $r$ be an extremal surface (see appendix A)
of the spatial geometry.
By comparison, $\ell$ increases monotonically as we move out from the
base of the geometry, insensitive to the formation of
extremal surfaces (or apparent horizons)
so that the identification of $\ell$ with $r$ is globally valid.

We can write the extrinsic curvature in the form consistent with
spherical symmetry

$$K_{ab}= n_a n_bK_{\cal L} + (g_{ab}-n_a n_b)K_R\,,\eqno(2.4)$$
where $K_{\cal L}$ and $K_R$ are two spatial scalars and
$n^a$ is the outward pointing unit normal to the two-sphere of fixed $r$,
$n^a=({\cal L}^{-1},0,0)$.
With respect to the proper timelike normal derivative
($N=1$ and $N^r=0$),
$K_{ab}= \dot g_{ab}/2$, so that
$K_{\cal L}=\dot {\cal L}/{\cal L}$ and $K_R=\dot R/R$.
$K_{\cal L}$ is also proportional to the acceleration
of a radial spacelike geodesic curve on the initial data surface.

The quadratic in $K_{ab}$ appearing
in the Hamiltonian constraint can be expressed in
terms of $K_{\cal L}$ and $K_R$.
The constraints are now given by

$$K_R\left[K_R+2K_{\cal L}\right]-
{1\over R^2}\Big[2 \left(R R^\prime \right)^\prime -R^{\prime 2}
-1 \Big]=8\pi \rho\eqno(2.5)$$
and

$$K_R^\prime + {R^\prime \over R}(K_R-K_{\cal L})=4\pi J\,,\eqno(2.6)$$
where we define the scalar $J=J\cdot n$.
All but the radial component of the current three-vector $J$ vanish.
The only non-vanishing momentum constraint is the projection onto the
radial direction.

In any realistic model, matter will be
modelled by a field theory. $\rho$ and $J$
will then be cast as functionals of the fields and their momenta.
However, for our purposes we will
suppose that we are given two functions $\rho(\ell)$
and $J(\ell)$ on some compact support, say $[0,\ell_0]$.
An important fact we will discuss in detail below is that
a solution which is both asymptotically flat and non-singular will not
exist for every specification of $\rho(\ell)$ and $J(\ell)$. This
might happen if an excessive energy, in a sense which will be
defined more precisely in subsequent papers, is concentrated within
a confined region [20,21].

We stress that
our specification of the sources possesses a spatial diffeomorphism
invariant meaning. This should be contrasted with
the provision of $\rho$ or $J$ as functions
of the flat background coordinate in conformal coordinates, $r {\cal L}=R$,
with respect to which the line element assumes the conformally flat
form

$$ds^2={\cal L}^2(dr^2+r^2 d\Omega^2)\,.\eqno(2.7)$$
Like the proper radial identification, this system is
globally valid.  The disadvantage is the
unphysical nature of the background spatial geometry.
Even the simple constant density star is not without its
subtleties in this gauge despite the fact that the constant density is
a spatial diffeomorphism invariant.
The reason is that the dimensions of the
physical support of the star is determined in terms of its
coordinate dimensions with respect to the flat background
only after we have solved the constraint.

In conformal gauge, an appropriate conformal scaling of $\rho$
is often introduced in order to guarantee existence of a solution to
the constraints[3].
The result is that one appears to be able to sidestep
the very singularities we take pains to focus on. While this is fine when
one is only interested in existence, simply consigning boundary points on
the configuration space to infinity does not help to clarify the
physics which underlies the occurence of singularities.

To be fair there is no procedure for solving the initial value problem
which is entirely satisfactory.
Even though the specification of $\rho$ as a
function of $\ell$ does possess a spatial diffeomorphism
invariant significance, we have no quantitative notion of the
proper volume it occupies or, indeed, if such a $\rho$ can be
even consistently specified until we solve the constraints.
In the former case, we could, of course, treat $V$ itself as
our spatial coordinate. This would correspond to the identification
$4\pi R^2 {\cal L}=1$. The constraints then provide an
equation for $R$ (and thus trivially also for
${\cal L}$). However, the benefit we gain is offset by the increased
non-linearity of the equations.

\vskip1pc
\noindent{2.2 Boundary Conditions}
\vskip1pc
We are interested in geometries which possess a single
asymptotically flat region. It is then appropriate
to require that the geometry be closed at one end, $\ell=0$:

$$R(0)=0\eqno(2.8a)\,.$$
In this way, we exclude the possibility that
the geometry possess a wormhole to another asymptotically flat region
or that it does something like degenerate into an infinite cylinder
at this end.  Local flatness of the metric at this
base point also requires that

$$R^\prime(0)=1\,.\eqno(2.8b)$$
A remarkable feature of the constraints is that once we
demand that the geometry be regular at its base point, this
boundary condition is automatically implemented when the constraints
are satisfied. The only boundary condition we need
to impose on $R$ is (2.8a).
The technical reason for this is the singularity of
the Hamiltonian constraint Eq.(2.5)  as a second order ODE. This is
obvious if we rewrite the constraint in the form

$$R R^{\prime\prime} =
{1\over 2}(1-R^{\prime2}) +
{R^2\over2} K_R[K_R+2K_{\cal L}]-
4\pi R^2\rho\,.\eqno(2.5^\prime)$$
The right hand side is regular if $R$ is and $K_{ab}$ blows up
no faster than $R^{-1}$. Because $R$ now multiples the second
derivative the equation must be singular at $R=0$. Once we impose
the boundary condition (2.8a), however,
the requirement that $R^{\prime\prime}$ also be finite
enforces Eq.(2.8b) (by convention, we choose the positive sign) and in
turn, $R^{\prime\prime}(0)=0$.
For a given $\rho(\ell)$ and $J(\ell)$,
a non-singular solution of the constraint will be unique
if it is regular at $\ell=0$.

In particular, we will also see that the single boundary condition
(2.8a) is  sufficient to guarantee that spacetime be asymptotically
flat, $R\to \ell$ as $\ell\to\infty$ provided
the sources are distributed on a compact support (or fall off
appropriately) and provided the geometry is non-singular.

As an illustration of what might go wrong if $R^\prime(0)
\ne 1$, let us compute the three-scalar curvature for
the spatial geometry described by the line element,
$ds^2=d\ell^2+a^2 \ell^2 d\Omega^2$ where $a$ is some positive constant.
In this geometry, $R^\prime(0)=a$.
If $a\ne 1$, the geometry suffers a conical singularity at the origin
associated with the solid angle deficit, $\Delta\Omega = 4\pi (1- a^2)$.
This manifests itself in the divergence of ${\cal R}$ given by

$${\cal R}={1\over 2\pi a^2 \ell^2}\Delta \Omega\,,$$
as the origin is approached.
The sign of ${\cal R}$ depends on the sign of $a-1$. It
is positive when $a>1$ (a solid angle surplus)
and negative when $a<1$ (a solid angle deficit).
Unlike a two-dimensional cone which is flat
away from its apex (${\cal R}=0$ when $\ell \ne 0$),
the conical singularity we are considering
has a long range field associated with it.
In fact, the falloff in ${\cal R}$ is so slow that
the space is not even asymptotically flat.
This is the generic behavior associated with
a conical singularity.
Two-dimensional conical structures are exceptional in this regard.

If both $J=0$ and $K_{ab}=0$ the hamiltonian constraint gives us that
${\cal R}$ will be finite when $\rho$ is.
The constraints therefore forbid simple conical singularies
(finite $R^\prime \ne 1$) under these conditions.\footnote * {One could
consider a distribution of $\rho$ which diverges like
$\ell^{-2}$ at the origin so that its integral over the spatial volume
in the neighborhood of $\ell=0$ is finite.}
They do, however, admit more serious cusp singularities (infinite $R^\prime$)
with a divergence in the traceless component of ${\cal R}_{ab}$.
If $J\ne0$, however, the constraints do not necessarily imply that
${\cal R}$ is finite. This is because  a divergence in ${\cal R}$ can be
balanced by a divergence in $K_{ab}$.
However, what is true is that $R^\prime$
will always diverge at the singularity so that conical singularities
cannot occur. The formation of singularities will be discussed
in sect.5 and in greater detail in II ($J=0$)[20] and
III($J\ne0$)[21].

\vskip1pc
\noindent{2.3 The Constraints in terms of the Optical Scalars}
\vskip1pc

A remarkable feature of the constraints
when the spacetime geometry is spherically symmetric is that the
constraint equations, (2.5) and (2.6) can be expressed in
a symmetrical form with respect to the optical scalars, defined
in terms of the divergence of the future pointing and past pointing
outward radially directed light rays on the
spherical surface of fixed proper radius. In appendix A, we show that

$$\Theta_\pm = {2\over R}(R^\prime \pm RK_R)\,,\eqno(2.9a,b)$$
{\it i.e} $\Theta_+(\Theta_-)$ is the
tangential projection of the sum (difference) of the
metric connection and the extrinsic curvature tensor.
In addition, $\Theta_+$ and $\Theta_-$ are canonically conjugate
variables. In the quantum theory, the $\Theta_+$ representation
appears to provide a very simple characterization of the states
which correspond to configurations without apparent horizons  of the form,
$\Psi(\Theta_+)=0,$ if $\Theta_+\le 0$.

We can invert the defining equations (2.9a) and (2.9b) in favor of
the tangentially projected two-extrinsic and three-extrinsic
curvatures,

$$\eqalign{ R^\prime =& {R\over 4}(\Theta_+ +\Theta_-)\,,\cr
            R K_R    =& {R\over 4}(\Theta_+ - \Theta_-)\,.\cr}
            \eqno(2.10a,b)$$
It is straightforward now to demonstrate that by
adding and subtracting an appropriate linear combination
of the constraints, Eqs.(2.5) and (2.6), we obtain two equivalent
constraints ($\omega_\pm = R\Theta_\pm $)

$$\eqalign{(\omega_+ )^\prime =&
-8\pi R(\rho-J) -{1\over 4R} (\omega_+\omega_--4) +\omega_+ K_{\cal L}\cr
(\omega_-)^\prime =&
-8\pi R(\rho+J) -{1\over 4R} (\omega_+\omega_--4) -\omega_- K_{\cal L}
\,.\cr}\eqno(2.11a,b)$$
We note that Eq.(2.11b) obtains from Eq.(2.11a) under time
reversal, $J\to-J$ and $K_{ab}\to-K_{ab}$.\footnote *
{A spacetime argument is presented in appendix B which
makes the existence of two two such equations more obvious.}
In this form, the two constraints are linear in the `momentum'. In
this sense they are the natural `square roots' of the Hamiltonian.
Note, however, that $K_{\cal L}$ appears on the RHS of Eqs.(2.11).
The most natural way to treat $K_{\cal L}$, in this context,
is as an independent initial datum specifying some
extrinsic time foliation.\footnote * {This is not, however, the
usual way to fix such a foliation, which generically will be
some functional relation of the form, $F(K_{\cal L},K_R)=0$.}
We note that these equations are simpler than the
equations written down by Malec and \'O Murchadha
who treat the trace of $K_{ab}$, instead of
$K_{\cal L}$, as the independent foliation
datum [15].

If the geometry is locally flat at the origin, so that
Eqs.(2.8a) and (2.8b) hold, and the tangential projection of the
extrinsic curvature diverges at the origin slower than $R^{-1}$
then
$$\omega_+(0)=2=\omega_-(0)\,,\eqno(2.12a,b)$$
If the geometry is asymptotic flat, in addition,
$$\lim_{R\to\infty}\omega_+=2=\lim_{R\to\infty}\omega_-\,.\eqno(2.13)$$

\vfill\eject
\noindent{\bf 3 FOLIATIONS AND SOLUTIONS OF THE
MOMENTUM CONSTRAINT}
\vskip1pc

To fix the foliation of spacetime, we will freeze some homogeneous
linear combination of the extrinsic curvature scalars. Such a choice is
natural because the momentum constraint is itself linear in $K_{ab}$.
Any non-linearity or inhomogeneity in the gauge condition would
destroy the linear scaling of $K_{ab}$ with $J$.

More specifically, let us consider the one-parameter family of
foliations,

$$ K_{\cal L}+\alpha  K_R=0\,.\eqno(3.1)$$
We do not consider gauges involving higher
spatial derivatives of the extrinsic curvature scalars.
The foliation (3.1), whenever valid, fixes one
(linear combination) of the two geometrical momenta at each point.
The remaining one is determined completely in terms
of the intrinsic geometrical and matter variables
by solving the momentum constraint. This reads

$$K_R^\prime + (1+\alpha){R^\prime\over R}
 K_R =4\pi J\,,\eqno(3.2)$$
and is exactly solvable. The solution which is
regular at the origin is
$$ K_R = {4\pi\over R^{1+\alpha}}\int_0^{\ell} d\ell R^{1+\alpha}
J\,.\eqno(3.3)$$
$K_{ab}$ vanishes if $J=0$ everywhere. In particular,
the foliation of Minkowski space by any one of these gauges is
the standard flat slicing.  If $K_R$ vanishes at any point, then so
also does
$K_{\cal L}$ so that $K_{ab}=0$ there.
We note that the boundary condition (2.8a) at the origin
implies that

$$K_R(\ell)\sim 4\pi {J(0)\over 2+\alpha}\ell\,,$$
in the neighborhood of $\ell=0$.
In particular, $K_R(0)=0$ and therefore so also we have that
$K_{ab} = 0$ at the origin.

The slowest acceptable falloff of the extrinsic curvature
in an asymptotically flat geometry must be faster than
$R^{-3/2}$ [22].
If $\alpha<0.5$ the solution of Eq.(3.2) is inconsistent with
asymptotic flatness of the metric in a spatially open geometry.

If $\alpha>0.5$ and is finite,
the gauge is globally  valid on any regular spatial
geometry regardless of the support of the initial data, or
the presence of extremal or trapped surfaces.

It is odd, therefore,
that the simplest choice of gauge in the spherically symmetric
context appears to be the polar gauge \footnote * {What might appear
to be the other `natural' possibility
$$K_{\cal L}=0$$
corresponds to $\alpha=0$ and therefore
does not  yield a satisfactory falloff.}

$$K_R=0\,,\eqno(3.4)$$
which corresponds to $\alpha\to\infty$ in the
parametrization (3.1).\footnote {$\dagger$}
{In phase space, this is expressed as
$\Pi_{\cal L}=0$ where $\Pi_{\cal L}$ is the momentum canonically conjugate
to ${\cal L}$.}
What is most alluring about this gauge in a
spherically symmetric geometry is that when $K_R=0$, the dependence on
$K_{ab}$ drops out of the Hamiltonian constraint (2.5)
which then mimics its
form in a momentarily static configuration (MSC) $K_{ab}=0$.
The Hamiltonian constraint can then be solved independently of the
value assumed by the unfixed extrinsic curvature scalar $K_{\cal L}$.
One particular peculiarity of this gauge is that apparent horizons
show up as extremal two-surfaces of the intrinsic spatial geometry
(see appendix A). Remarkably, in fact, the physical content of the model
gets encoded completely in
this geometry.

Where then is the snag?
To see what price we have to pay, let us examine the
solution of the
momentum constraint (2.6)
$$ K_{\cal L}=-4\pi {R\over R^\prime}J\,.\eqno(3.5)$$
$ K_{\cal L}$ is determined locally in terms of the source.
This contrasts dramatically with the solution
Eq.(3.3) when $\alpha$ is finite where $ K_{\cal L}$ is determined
non-locally in terms of $J$. As $\alpha$ tends
to infinity, the differential equation (3.2)
becomes singular. The support of the integrand
appearing in the solution collapses
in this limit and we recover Eq.(3.5).

Technically, this is because no derivative of $ K_{\cal L}$
appears in the constraint. The consequence, however,
is that the gauge will break down whenever $R^\prime$ vanishes
on the support of $J$. The vanishing of
$R^\prime$ signals the development of an extremal two-surface
in the spatial geometry so that the gauge breaks
down whenever a current flows across an
apparent horizon.

The foliation gauge condition should also fix the lapse.
It is easily demonstrated that
if $R^\prime=0$ anywhere on the support of $\rho$, the lapse
collapses $N(\ell)\to 0$ in polar gauge --- another manifestation of the
breakdown of the gauge. Polar gauge clearly does not provide a
useful description of the physics
in the strong field regime (inside matter)
we are interested in.

Outside the support of $J$, Eq.(3.5) implies that
$ K_{\cal L}=0$ so that $ K_{ab}=0$.
This means that the slicing of spacetime defined by polar gauge
coincides with the sequence of level surfaces of the timelike killing vector
in this region.
Vacuum spacetime therefore appears `static' in polar gauge.
This is the optimal exterior form of the metric.
In fact, as we can see, polar gauge is the unique member of
the one-parameter family possessing this property.
The only obvious shortcoming of finite
$\alpha$-gauges is that they
do not provide a static slicing outside the support of the matter.

It is also clear, however, that a static description of
spacetime can be approximated arbitrarily
closely outside the support of matter by letting $\alpha$ be
appropriately large. This suggests the possibility of
constructing a gauge which displays
the exterior behavior of polar gauge, while
at the same time side-stepping its interior shortcomings.
What we can do is to
admit a space dependent parameter $\alpha(\ell)$
in Eq.(3.1) which tends to infinity outside matter.
Let $\alpha(\ell)=\alpha+\beta(\ell)$, such that
$\lim_{\ell\to\infty}\beta(\ell)=\infty$. Then
$$ K_R = -{4\pi\over R^{1+\alpha}F}\int_0^{\ell}
d\ell R^{1+\alpha} F J\,,\eqno(3.3^\prime)$$
where $F(\beta,\ell)= {\rm exp}\,
(\int_0^\ell d\ell \beta R'/R )$.
There does not appear to be  any
gauge (intrinsic or extrinsic) which is not
tuned artificially by hand which will provide a static description of
spacetime outside matter.

The gauge (3.4) is clearly not the only
linear combination of the geometric momenta
in which the Hamiltonian constraint (2.5) mimics the MSC form.
The  gauge $2K_{\cal L}+ K_R =0$ will also do the job.
While this does not appear to
suffer from the pathologies of polar gauge, it suffers
from the shortcoming of producing a slow falloff $\sim R^{-3/2} $
in $K_{ab}$ outside the support of the current. This
complicates the asymptotic analysis of the field. As we will see,
the conventional expression for the ADM mass is no longer valid.

The existence of two gauges mimicing a MSC is a consequence
of the Lorentz signature $(-,+)$ of (the metric part of)
the supermetric  which permits the
term quadratic in the metric momenta to factorize.
These two gauges define the null directions in superspace
with respect to the supermetric.
With respect to a foliation defined by any other linear combination of
$K_{\cal L}$ and $K_R$, extrinsic curvature will show up in
Eq.(2.5).

The two MSC mimicking gauges with $\alpha=0.5$ and $\alpha=\infty$
define the light cone of the superspace. The admissible
gauges constructed using linear combinations of $K_{\cal L}$ and $K_R$
therefore correspond to tangent vectors lying strictly
inside the  light-cone of the superspace metric[16].
The trajectory in the configuration space therefore
takes place along spacelike directions.
This suggests a special role for the light-cone in
this mini-superspace.

The maximal slicing condition
$K=K_{\cal L}+2K_R=0$ corresponds to $\alpha=2$. This is the gauge
which most readily facilitates the analysis of the
constraints in York's conformal approach to the full theory and
remains the most popular choice among the more formally
inclined workers in the field. In the spherically
symmetric asymptotically flat context, however, this
is not such a convincing criterion. Any valid $\alpha$ would appear
to offer the same reasonable compromise
between acceptable asymptotic falloff and
non-singular behavior in the interior.
The remarkable nature of maximal slicing will, however,
become evident in Sect.5 within the framework of the optical scalars
introduced in Sect.2.3 when we specialize to initial data
satisfying the dominant energy condition.
This is not at all obvious in the context of the
metric variables we have been exploiting in this section.

We note that when $\alpha=1$,
the integral appearing in Eq.(3.3) is simply the
proper volume integral of the radial current
scalar $J$ in the interior of $\ell$. As we will see in paper III,
various results simplify dramatically in this gauge.\footnote *
{This gauge, like polar gauge, has a particularly simple
phase space representation, corresponding as it does to
the vanishing of the momentum canonically conjugate to $R$.}

We note that the solution of the momentum constraint requires us to
integrate out from $\ell=0$.
$ K_R R^{1+\alpha}$ tends asymptotically to the constant value
$$\int_0^\infty d\ell R^{1+\alpha} J\eqno(3.6)$$
determined by the current flow.
Thus outside the support of the field flow $K_R$
is proportional to $R^{-(1+\alpha)}$.
For example, when $\alpha=2$,  $ K_{ab}$ is, up to a constant,
the unique spherically symmetric transverse-traceless tensor on
$R^3$ (see [23]).

It might appear that we could just as well have chosen to integrate
Eq.(3.2) in from infinity and to have concluded that
outside the support of the field flow, the solution is always
$ K_R =0$ and that the foliation of spacetime is static.
The difficulty with this is that
the resulting solution will be singular at the origin
unless the current is fine-tuned appropriately.

Once the momentum constraint has been solved, we substitute
(3.1) and (3.3) into Eq.(2.5) and solve for $R(\ell)$
subject to the boundary conditions, (2.8). We defer the
details to papers II and III.

\vfill\eject

\noindent{\bf 4 THE QUASI-LOCAL MASS}
\vskip1pc

An important feature of the constraints when the geometry is
spherically symmetric is that they possess a first integral which
permits the definition of a quasi-local mass (QLM), $m(\ell)$,
over a sphere of fixed proper radius which can be
expressed as a volume integral over the
sources contained within that sphere.\footnote *
{The `integrability' of the system should not
be surprising once we identify the mechanical analog of
Eq.(2.5) corresponding to the identification of $\ell$ with time.
A generic two or higher dimensional model will
not be integrable in this way. There has been a flurry of research
recently on integrable `one' and
`two' dimensional models in general relativity [24].}
To motivate its definition, as well as to make a few observations about
asymptotic behavior, let us first consider the momentarily static
data, $K_{ab}=0$. We define

$$m_{K=0}={R\over 2}
\left(1-\left(R^\prime\right)^2\right)\,.\eqno(4.1)$$
This should be viewed as a surface integral over the sphere
of proper radius $\ell$ of a spherically symmetric scalar function.
When $K_{ab}=0$, it is simple to show that the
Hamiltonian constraint can be cast in the form

$$m_{K=0}^\prime = 4\pi R^2 R^\prime \rho\,,\eqno(4.2)$$
where $m$ is given by Eq.(4.1) for all values of $\ell$.
In particular, outside the
support of matter $m_{K=0}^\prime=0$ so that $m$ assumes a
constant value, $m_\infty$ say. This is the ADM mass.
If we implement regularity at $\ell=0$, Eq.(2.8), we obtain

$$m_{K=0} = 4\pi \int_0^\ell\, d\ell R^2 R^\prime\rho\,.\eqno(4.3)$$
Asymptotically, we can now rewrite Eq.(4.1)

$$R^{\prime2} = 1- {2 m_\infty\over R}\,.\eqno(4.4)$$
Thus, as $R\to\infty$, $R\sim\ell$ to leading order.
The ADM mass in encoded in the next to leading order,
$R \sim \ell - m_\infty\ln \ell$.
This is turn permits us to identify a simpler asymptotic
expression for $m_\infty$:

$$m_\infty =\lim_{\ell\to\infty} R(1-R^\prime)\,.\eqno(4.5)$$

If $K_{ab}$ does not vanish, the naive generalization is
to replace the quadratic $R^{\prime2}$ by the
square of the spacetime covariant derivative,

$$m := {R\over 2}
\left(1-\nabla_\nu R \nabla^\nu R\right)\,.\eqno(4.6)$$
Using the fact that $K_R = \dot R/ R$, this
yields the expression

$$m = {R^3 K_R^2\over 2}+{R\over 2}\left(1-
(R^\prime)^2\right)\,,\eqno(4.7)$$
which depends, as before, only on initial data.
To determine the form of the first integral of the
constraints analogous to Eq.(4.2), we integrate the
the Hamiltonian constraint up to $\ell$ to obtain

$${R\over 2}\left(1-\left(R^\prime\right)^2\right)
=4\pi \int_0^{\ell} d\ell \rho R^2 R^\prime
-{1\over 2 }\int_0^{\ell}
d\ell K_R\left[K_R+2K_{\cal L}\right]R^2 R^\prime
\,.\eqno(4.8)$$
As before, the LHS coincides asymptotically with $m_\infty$.
However, it does not coincide with $m_\infty$ outside the
support of matter.

Let us add the extrinsic curvature quadratic appearing in the
definition (4.7) to the LHS of Eq.(4.8). We can exploit the
momentum constraint (2.6) to eliminate
$K_{\cal L}$ appearing in the integral on the RHS in favor of $K_R$ and $J$.
Thus, modulo the constraints, $m$ satisfies

$$m=4\pi \int_0^{\ell} d\ell R^2\left[\rho R^\prime
+ J R K_R \right]\,.\eqno(4.9)$$
The RHS is the integral of the scalar,
$\mu := \rho R^\prime  + J R K_R $,
over the volume bounded by the surface at proper radius $\ell$
($dV=4\pi R^2 d\ell$)(see [19]). It is clear that, outside the support of
matter,  $m$ is a constant which we again identify as the ADM mass,
$m_\infty$. \footnote *
{We will see below that the LHS of (4.8), like $m$, is positive
everywhere in any regular geometry
when spacetime is foliated by an $\alpha$ gauge.
In contrast to $m$, it is even positive everywhere
in Euclidean relativity.}

We note that if the extrinsic curvature scalar
$K_R$ (as well as $K_{\cal L}$ )
tends asymptotically to zero faster than $R^{-3/2}$,
(4.7) reduces asymptotically to the same form (4.5) as Eq.(4.1).
However, when $\alpha=0.5$ the asymptotic form of the
surface integral Eq.(4.7) is not (4.1).
To see this, let us examine the asymptotic
`dependence' of the extrinsic curvature
contribution to $m$ on the parameter $\alpha$.
We note that ${R^3 K_R^2\over 2}\sim R^{1-2\alpha}$
tends to a constant if $\alpha=0.5$ and
diverges if $\alpha<0.5$. The latter
possibility was rejected in sect.3
because it was inconsistent with asymptotically flat boundary conditions.
We can also see that such a foliation yields an
asymptotically divergent QLM. In particular, it does not
coincide asymptotically with $m_\infty$.

We noted in sect.2 that if the geometry is non-singular, and the sources
have compact support  then regularity at the origin
is sufficient to force asymptotic flatness. This point is
clarified using the first integral of the constraint encoded in the
definition of the QLM. From Eq.(4.9) it is clear that if
both $\rho$ and $J$ are compactly supported and $R^\prime$ and
$K_R$ remain finite, the QLM will be a finite constant outside the
support of matter. We also noted in the
last paragraph that if $K_R$ falls off fast enough
then (4.7) also reproduces Eq.(4.4) so that
$R$ approaches $\ell$ in the same way as it does for momentarily
static data.

Only one linear combination of the constraints features in the
derivation of (4.9). It proves
extremely useful to exploit this first integral of
the constraints, implementing regularity at the origin,
in place of one or the other of the constraints.
In practice, we replace Eq.(2.5) by (4.9) (with
$m$ defined by (4.7)).
If, in turn, we suppose that spacetime is foliated
by an $\alpha$ gauge, then we can solve Eq.(2.6)
for $K_R$ in terms of $J$ obtaining the
expression given by Eq.(3.3).

Note that we have eliminated $K_{\cal L}$ in going from Eq.(4.8)
to Eqs.(4.7)/(4.9) without any recourse to a foliation gauge condition.
Two properties of the constraints have conspired to
yield the simple form for $\mu$ as a local scalar.
The first is that $K_{\cal L}$ appears linearly in Eq.(2.5)
and therefore linearly in Eq.(4.8). The second is that
it appears undifferentiated in Eq.(2.6).
In both regards it is unlike $K_R$. There is clearly a
conspiracy involving both constraints permitting the QLM to be expressed
in the simple form Eq.(4.9).

It is extremely useful to cast the QLM in terms of
the optical scalars, $\Theta_+$ and $\Theta_-$. We get

$$
m= {R\over 2}\Big(1 -{1\over 4}\omega_+\omega_-\Big)\,.\eqno(4.10)$$
The quasilocal mass $m$ is seen to be just the Hawking mass [25].
With respect to these variables, Eq.(4.9) assumes the form

$$
m= \pi \int_0^{\ell} d\ell R^2\left[\rho (\omega_+ +\omega_-)
+ J (\omega_+-\omega_-)\right]\,,\eqno(4.11)$$
or

$$m= \pi \int_0^{\ell} d\ell R^2\left[(\rho+J) \omega_+
+ (\rho- J)\omega_-\right]\,.\eqno(4.11^\prime)$$
The ADM mass is a spacetime diffeomorphism invariant. In particular,
it is independent of the foliation of spacetime on
which it is constructed. The definition of
the quasilocal mass either in terms of metric variables, Eq.(4.6)
or in terms of the optical scalars shows that
it is a spacetime scalar though its value does depend on the
foliation for finite values of $\ell$. For each value of $\ell$,
Eq.(4.9) provides a quasi-local observable of the classical theory.
In addition, these observables are non-trivial. For whereas
$m_\infty$ as the effective Hamiltonian is trivially
conserved, the observables defined by the QLM  are not.

We note that, in general, the spacetime
covariant derivative of $m$ can be cast in the form [9]

$$\nabla_\mu m = {R^2\over 2} G^{\alpha\beta} \epsilon_{\alpha\mu}
\epsilon_{\beta\nu} \nabla^\nu R \,,\eqno(4.12)$$
where the notation we use has been defined in appendix B.
The Einstein equations can now be exploited to recover
Eq.(4.9) on projecting (4.12) along the radial direction.
The evolution of $m$ along the (timelike) normal to the
hypersurface, $t^\mu$,
is obtained by projecting (4.12) onto $t^\mu$. Note that the radial
pressure will occur on the RHS.

\vskip2pc

\noindent{\bf 5 BOUNDS ON THE PHASE SPACE VARIABLES
AND THE POSITIVITY OF THE QLM }
\vskip1pc

The most important property of the QLM
is its positivity everywhere in any regular
spherically symmetrical geometry. In this section we will
demonstrate how this positivity arises as a consequence
of bounds on the phase space variables.

\vskip1pc
\noindent{5.1 Positivity of $m$:
The dominant energy condition, and a bound on the
product of the Optical Scalars }
\vskip1pc
Let us suppose that the material energy current four-vector
is timelike so that it satisfies the dominant energy condition (DEC) [10],

$$\rho \ge \sqrt{J^a J_a} \,.\eqno(5.1)$$
Suppose also that the constraints (2.5) and (2.6), or
alternatively (2.11a) and (2.11b), are satisfied. Then $m$
is positive everywhere, independently of how we foliate spacetime,
if the spatial geometry is regular everywhere. Because $m$ coincides
with $m_\infty$ at infinity, this provides us with
a generalization of the positivity of the ADM mass.

Because the sources appear explicitly in the manifestly
positive combinations, $\rho\pm J$, when the constraints are
cast in terms of $\omega_+$ and $\omega_-$, it suggests that
these are the more appropriate variables to use when the
DEC can be exploited.

Recall the definition of the QLM in terms of the
optical scalars, Eq.(4.10). It is clear that the positivity of
the QLM is entirely equivalent to the statement

$$\omega_+\omega_- \le 4\,.\eqno(5.2)$$
This inequality was first derived in [15] but, for completeness, we give a
derivation here. We note that we can exploit Eqs.(2.11a) and
(2.11b) to obtain

$$(\omega_+\omega_-)^\prime = -8\pi R\Big[(\omega_++\omega_-)\rho
- (\omega_+-\omega_-)J\Big] -
(\omega_++\omega_-)(\omega_+\omega_- - 4)\,.\eqno(5.3)$$
Note how $K_{\cal L}$ which appears in both Eqs.(2.11a) and
(2.11b) has dropped out of Eq.(5.3).
This equation is entirely equivalent to Eq.(4.11) with
$m$ defined by (4.10).
We note first of all that the product satisfies the
boundary conditions

$$\omega_+\omega_-(0)=4=\lim_{R\to\infty}\omega_+\omega_-\,,\eqno(2.12c)$$
on account of the boundary conditions,
Eqs.(2.12) at the origin and (2.13) at infinity,
if the geometry is asymptotically flat. In addition, it
must be finite everywhere in any regular geometry.
And, if the product is finite everywhere, it
must possess an interior critical point at
some finite value of $\ell$ if it is not constant.
At the critical point, the RHS of Eq.(5.3) must vanish.
Thus

$$\omega_+\omega_- - 4= -8\pi R\Big[\rho
-
{\omega_+-\omega_-\over \omega_++\omega_-} J \Big] \,,\eqno(5.4)$$
unless $\omega_+ + \omega_-=0$.
It is now clear that Eq.(5.2) is satisfied when $\rho\ge |J|$. For
if $\omega_+$ and $\omega_-$ possess different signs (which includes
the case where $\omega_+ + \omega_-=0$) then Eq.(5.2) is  obviously
satisfied. Therefore suppose that they possess the same sign. It is then
always true that

$$\Big|{\omega_+-\omega_-\over \omega_++\omega_-} \Big|\le 1\,.\eqno(5.5)$$
Thus the term appearing in square brackets in Eq.(5.4) is
manifestly positive whenever Eq.(5.1) is satisfied. This
establishes Eq.(5.2).

In those regions of the $(\omega_+,
\omega_-)$ plane where $\omega_+$ and $\omega_-$ possess
different signs so that $\omega_+\omega_-\le 0$,
not only is $m$ positive but, in addition,
$m\ge R/2$ regardless of whether the
constraints are satisfied, or that the energy is positive.
In particular, $m = R/2$  on the future and the
past apparent horizons.

We note that the absolute maximum of the product,
$\omega_+\omega_-$ obtains at
the boundary values $\ell=0$ and $\ell=\infty$ and
it is also the flat space value.
We note, however, the corresponding values of $m$ are
$m(0)=0$ and $\lim_{\ell\to\infty} m=m_\infty$.

\vskip1pc
\noindent{5.1.1 Negative QLM and the approach to singularities}
\vskip1pc
If $\omega_+\omega_->4$ anywhere, the geometry must possess a singularity.
How does this occur?

If we enter a region in which
$\omega_+\omega_->4$ with both $\omega_+$ and $\omega_-<0$, then
when the DEC is satisfied, Eq.(5.3) implies that

$$(\omega_+\omega_-)^\prime >0\,.\eqno(5.6)$$
The product $\omega_+\omega_-$ monotonically increases. Once
$m$ goes negative it decreases montonically.\footnote * {We will see in
II and III that it is nonetheless remains finite all the way to any
singularity.}  In particular, $m$ cannot recover positive values.
The barrier,
$\omega_+\omega_-=4$, with both $\omega_+$ and $\omega_-<0$ is therefore
semi-permeable.

In addition, when $\omega_+\omega_->4$, then

$${1\over 4}(\omega_+ +\omega_-) <-1\,,\eqno(5.7)$$
so that $R^\prime <-1$ and decreasing. Therefore, if the circumferential
radius is $R_0$ when $\omega_+\omega_-=4$ we know that the
solution must crash, {\it i.e.}, $R\to 0$ in a finite proper distance
which is less than $R_0$ from that point.

What would happen if instead we had a source which did not satisfy the DEC
so that we entered the region $\omega_+\omega_->4$ but
with both $\omega_+$ and $\omega_->0$? Let us further assume that the
source changed its nature so that in this region it did satisfy the
DEC. Instead of Eq.(5.6), we now have

$$(\omega_+\omega_-)^\prime < 0\,,\eqno(5.6^\prime)$$
which means that the solution is being pushed out of the region
$\omega_+\omega_->4$ and, instead of Eq.(5.7) we get
$${1\over 4}(\omega_+ +\omega_-) >1 \,,\eqno(5.7^\prime)$$
so that $R^\prime >1$. Hence there is no way that the
solution can now crash within the region $\omega_+\omega_->4$. Thus the
barrier  $\omega_+\omega_-=4,\omega_+,\omega_->0$ is also semi-permeable,
one can go from above down but not up from below so long as the DEC holds.

\vskip1pc
\noindent{5.2 Positivity of $m$:
$\alpha$-gauges, the weak energy condition and a bound on $R^{\prime2}$}
\vskip1pc

A remarkable feature of foliations of
spacetime by the $\alpha$ parameterized gauges is that
it is possible to (1) relax Eq.(5.1) to the weak
energy condition (WEC), {\it viz.}, $\rho\ge 0$ and,
(2) omit the momentum constraint yet still
establish the positivity of $m$ everywhere.

The proof is again very simple. However,
there is no advantage to be gained by exploiting the optical scalars.
We return to the definition of $m$ given in terms of
the metric variables, Eq.(4.7). It is clear that
$m\ge 0$ whenever $(R^\prime)^2\le 1$.
We need therefore only show that $(R^\prime)^2\le 1$ under the conditions
of the hypothesis and we are done.

We first note that $R^\prime$ must be bounded
in any regular geometry. Because
$R^\prime=1$ both at the origin and at infinity,
$R^\prime$ must possess some interior critical point.
This will occur when $R^{\prime\prime}=0$.
The Hamiltonian constraint (2.5) then implies that
$$(R^\prime)^2 =1-8\pi \rho R^2
+K_R\left[K_R+2K_{\cal L}\right]R^2
\eqno(5.8)$$
at such a points.
In a MSC,  if $\rho \ge 0$ at the critical point
then it is certainly always true that
$$(R^\prime)^2 \le 1\,.\eqno(5.9)$$
Unfortunately, if $K_{ab}\ne 0$ this will not generally be
true unless
$K_R\left[K_R+2K_{\cal L}\right]\le 0$ at this point.
There is, unfortunately, no gauge invariant reason why this should hold.
If, however, spacetime is foliated by any gauge
such that Eq.(3.1) holds at least at the critical points of $R^\prime$, then
the third term on the right hand side of Eq.(5.1) is given by

$$K_R\left[K_R+2K_{\cal L}\right]=(1-2\alpha) K_R^2\,,$$
which is negative if $\alpha> 0.5$. This
is just the condition defining a globally valid $\alpha$-gauge.
This completes the proof that $(R^\prime)^2\le 1$.

A few comments on the proof:

We note that the inequality (5.9) is stronger
than the inequality (5.2). It is clear that what the
lemma we have proved here is
a stronger statement than the positivity of the QLM.
For it is possible that $R^{\prime2}>1$ but $m>0$
(see  fig.(8.1)). In paper III we will demonstrate
that whereas the converse of the
positivity of the QLM is false ({\it viz.}, $m$ may be
positive everywhere but the geometry singular) the
converse of what we have proved is also true: the geometry is regular
if and only if $-1<R^\prime\le 1$.

Because the momentum constraint
has not featured in this proof, unlike the
first proof no control is
necessary over the material current such as that implied by
the DEC. However, in the same way that Eq.(4.11) involves only one linear
combination of the constraints, so does Eq.(5.3). Thus neither proof
requires the full constraints.

We note that when $\ell\to\infty$ we again recover the
positivity of the ADM mass. The proof
is interesting because, unlike the general proof [26],
it does not require the dominant energy condition to be satisfied.

Let us suppose that $m<0$ somewhere, so that
$R^{\prime2}>1$ at that point. However, if this is the case,
then Eq.(2.5) (or (2.5$^\prime$)) implies that

$$R^{\prime\prime} <0\,,$$
so that $R^\prime$ is decreasing there. This can only occur if
$R^\prime<-1$. Therefore, if the circumferential
radius is $R_0$ when  $m(\ell_0)=0$,
then $R^\prime\le -1$. We know then that the
solution must crash, {\it i.e.} $R\to 0$ in a finite proper distance
which is less than or equal to $R_0$ from that point.

In fact, we do not even require that $m<0$. For, as
we remarked before, it is
possible to have $R^\prime<-1$ but $m>0$.
Indeed, it is possible that though $R^\prime$ decreases monotonically,
$m$ nonetheless remains positive. The catalog of possibilities
will be discussed in papers II and III.

We finally recall that a complete specification of the
gauge was not necessary in the positivity proof provided in
this section. However, if we had
not specified the gauge everywhere $R^\prime<-1$
was satisfied, we would not
have been able to claim that $R^\prime$ decreased monotonically
as we proceeded outwards.

\vskip1pc
\noindent{5.3 $\alpha$ gauges and
embedding in Euclidean $R^4$}
\vskip1pc
If the inequality $R^{\prime2}\le 1$ holds everywhere on the interval
$[0,\ell]$ it is clear that the interior geometry can always be embedded
as a hypersurface in flat Euclidean $R^4$.
Thus any regular spherically symmetric
asymptotically flat three-geometry consistent with the
Hamiltonian constraint in the gauge Eq.(3.1)
can be embedded in $R^4$. Later we will
encounter (strongly) singular solutions of the
constraints which cannot be thus embedded [20,21].\footnote
{$^\dagger$} {We note that the lowest dimension into which the
Schwarzschild solution can be embedded is $R^6$ [27]
consistent with the singularity of the geometry at $R=0$.}

More generally, whenever
$K^{ab}K_{ab} - K^2$ is positive, it is clear from
Eq.(2.1a) that the scalar curvature, ${\cal R}$,
is also positive when $\rho$ is.
Intuitively, one would expect a spatial geometry
with a positive ${\cal R}$ to be embedded more readily
in a low dimensional flat Euclidean space (in the best case,
as a hypersurface) than a generic geometry.
This is, of course, not true of the embedding of
such a geometry in Lorentzian flat $R^4$ which requires
that the initial data be trivial.

Analogous statements do not exist in Euclidean relativity
where the sign of the extrinsic curvature quadratic is reversed.
Indeed, in the Euclidean theory, $\rho$ does not even
possess a definite sign. If $\rho$ is of the form kinetic energy plus
potential energy, the sign of the kinetic term will
reverse in the Euclidean theory.
There is, therefore, no analogous positive quasi-local mass result
for instantons except when $K_{ab}=0$. Indeed,
$R^{\prime2}$ need not be bounded at finite values of $\ell$.

\vskip1pc
\noindent{5.3.1 A Universal Bound on $R$ by $\ell$}
\vskip1pc
We have seen that the only possible approach to a singularity is
through $R^\prime\le -1$. It is therefore always true that
$R^\prime\le 1$ on any slice defined by an $\alpha$-gauge. If this
inequality holds everywhere on the interval $[0,\ell]$, the proper radius
of the geometry will exceed its circumferential radius at $\ell$. This is
simply because then, $\ell-R =\int_0^\ell d\ell (1-R^\prime)\ge
0$. Equality only obtains when space is flat.

\vfill\eject
\noindent{\bf 6 BOUNDS ON $\omega_+$ AND $\omega_-$}
\vskip1pc

In general, the inequality (5.9) will not be valid,
even when the dominant energy condition is satisfied,
if the gauge is not of the form (3.1). The embedding
argument we have just considered in Sect.5.3 is not gauge invariant.

In this section we
will demonstrate that when Eq.(5.1) is satisfied, both
$\omega_+^2$ and $\omega_-^2$ are bounded when the
full constraints are satisfied. The nature of these bounds
is very different from that of the upper bound Eq.(5.2) we obtained on the
product $\omega_+\omega_-$.
The proof that such bounds exist proceeds, however, in exactly the
same way as the proof of (5.2). Before proceeding with the
proof, it is useful to recast the constraint equations
Eqs.(2.11a) and (2.11b) in a form which treats the trace, $K$,
rather than $K_{\cal L}$ as the independent extrinsic curvature scalar which
will be fixed by an appropriate gauge condition.
Eliminating $K_{\cal L}$ in favor of $K$ we get [15]

$$\eqalign{(\omega_+)^\prime=
&-8\pi R(\rho-J) -{1\over 4R}
\Big[2\omega^2_+ -4 - 4\omega_+ KR -\omega_+\omega_-\Big]\,,\cr
(\omega_-)^\prime=
&-8\pi R(\rho+J) -{1\over 4R}
\Big[2\omega^2_- -4 + 4\omega_- KR -\omega_+\omega_-\Big]\,.\cr}
\eqno(6.1a,b)$$
We will prove that

$$|\omega_\pm|\le |\kappa| +
\Big(|\kappa|^2+4\Big)^{1/2} := \Omega\,,\eqno(6.2a,b)$$
where $\kappa ={\rm Sup}\, R |K|$.
We again recall that both $\omega_+$ and $\omega_-$ satisfy the
boundary conditions Eqs.(2.12) at the origin and (2.13) at infinity
if the geometry is asymptotically flat. In addition, they must
be finite everywhere in any regular geometry.
And, if so, they each must possess an interior critical point at
some finite value of $\ell$ if not constant.
At the critical point, the RHS of Eq.(5.3) must vanish.
Thus

$${1\over 4R}\Big[2\omega^2_+ -4 - 4\omega_+ \kappa_+\Big]
= {1\over 4R}\omega_+\omega_- - 8\pi R(\rho-J) \,,$$
where $\kappa_+$ is the value of $R K$ at the critical point
of $\omega_+$.
Exploiting Eqs.(5.1) and (5.2) we have

$$\omega^2_+ - 2\omega_+ \kappa_+ - 4 \le 0,,\eqno(6.3)$$
It is now clear that

$$\kappa_+-
\Big(|\kappa_+|^2+4\Big)^{1/2}\le \omega_+\le
\kappa_+ +
\Big(|\kappa_+|^2+4\Big)^{1/2}\,.\eqno(6.4a)$$
Similarly

$$-\kappa_- -
\Big(|\kappa_-|^2+4\Big)^{1/2}\le \omega_-\le
-\kappa_- +
\Big(|\kappa_-|^2+4\Big)^{1/2}\,,\eqno(6.4b)$$
where $\kappa_-$ is the value assumed by $RK$ at
the critical point of $\omega_-$. Now both $|\kappa_+|$ and $|\kappa_-|$
are bounded by $\kappa$ which completes the proof of
Eqs.(6.2a) and (6.2b).
We note that we can obtain more transparent (though weaker)
inequalities, by further approximating $(|\kappa|^2 + 4)^{1/2}
\le |\kappa| + 2$, so that
$|\omega_\pm|\le 2 + 2|\kappa|$.

More stringent bounds on $\omega_+$ and $\omega_-$ can be extracted
from Eqs.(6.4a,b) by using $\kappa^*_+ = \max RK$ and $\kappa^*_- =
\max(-RK)$ to give
$$-\kappa^*_- -
\Big(|\kappa^*_-|^2+4\Big)^{1/2}\le \omega_+\le
\kappa^*_+ +
\Big(|\kappa^*_+|^2+4\Big)^{1/2}\,.\eqno(6.6a)$$
$$-\kappa^*_+ -
\Big(|\kappa^*_+|^2+4\Big)^{1/2}\le \omega_-\le
\kappa^*_- +
\Big(|\kappa^*_-|^2+4\Big)^{1/2}\,.\eqno(6.6b)$$
These expressions are useful in the special case where $K$ has a fixed
sign. Consider the case where $K \ge 0$. This gives $\kappa^*_- = 0$
and we get the bounds
$$-2 \le \omega_+\le
\kappa^*_+ +
\Big(|\kappa^*_+|^2+4\Big)^{1/2}\,,\eqno(6.7a)$$
$$-\kappa^*_+ -
\Big(|\kappa^*_+|^2+4\Big)^{1/2}\le \omega_-\le 2\,,\eqno(6.7a)$$
and in the case where $K \le 0$ we get
$$-\kappa^*_- -
\Big(|\kappa^*_-|^2+4\Big)^{1/2}\le \omega_+\le 2\,.\eqno(6.8a)$$
$$ -2\le \omega_- \le
\kappa^*_- +
\Big(|\kappa^*_-|^2+4\Big)^{1/2}\,.\eqno(6.8b)$$
Of course, if spacetime is foliated by an $\alpha$-gauge, then
we can do better still. If $J\ge (\le)0$, then $\omega_+\ge
(\le)\omega_-$.

We can also place a bound on the sum
and  difference of $\omega_+$ and $\omega_-$. We exploit Eqs.(6.2a \& b)
to get, for both the sum and difference

$$|R^{\prime}|, |RK_R|\le  {1\over 2}\Omega\,.\eqno(6.9)$$
When $K=0$, the former bound coincides
with the bound (5.7) in $\alpha$ gauges.
If $K\ne 0$, $R^{\prime2}$ is still bounded if $K$ is.
However, the corresponding spatial geometry will not
generally be embeddable in $R^4$.
The inequality on the difference has no analog
if the dominant energy condition is not satisfied.

It is clear that the bound on the sum can be improved. This
is because the bound on the product, Eq.(5.2), does not
permit $\omega_+$ and $\omega_-$ to simultaneously
saturate their upper and lower bounds. We find

$$|R^{\prime}|\le  {1\over 4}\Big(\Omega + {2\over \Omega}\Big)\,.$$
This is most easily checked using the
the graphical representation provided in fig.(8.1).

The inequalities, Eqs.(6.2a) and (6.2b)
come cast naturally in terms of $K$.
Despite the fact that the pair of equations,
Eqs.(2.11a) and (2.11b), superficially appear simpler than
(6.1a) and (6.1b) the latter provide
the more natural presentation of the constraints.

A privileged role appears to be played by the maximal slicing
of spacetime. Now $|\omega_\pm|\le 2$. The bounds Eqs.(6.2a) and (6.2b)
now imply the bound (5.2). It might appear that
when $\alpha\ne 2$ the bounds on $\omega_+$ and
$\omega_-$ are not so useful
appearing as they do to involve ${\rm Sup}\, R|K|$ explicitly.
One can show, however, by bootstrapping on these inequalities that
that the numerical bound which independent of $|K|$,
$|\omega_\pm|\le 2/\sqrt{1 - |2-\alpha|}$, can be
established in the neighborhood of $\alpha=2$ [21].
We now independently possess the bound $R'^2\le 1$ to establish (5.2).

\vskip1pc
\noindent{\bf 7 NEGATIVE BINDING ENERGY}
\vskip1pc

Once we fix a foliation, a simple measure of the material energy
content of the system is the (non-conserved) quantity

$$M=4\pi\int_0^{\ell} d\ell R^2 \, \rho\,, \eqno(7.1)$$
also termed the `bare mass' by ADM [28].
$M$ is a spatial scalar. It is positive and monotonically
increases with $\ell$ when $\rho$ is positive.
It does, however, depend on the foliation.
It is the QLM which we can think of as the sum of $M$ and a
deficit which we tentatively identify with the
gravitational binding energy $E_B$ associated with the
sources in its interior

$$m=M+E_B\,,\eqno(7.2)$$
which is independent of the foliation. Its value at infinity is
also conserved.

On the other hand, even
though the QLM is positive everywhere when the dominant energy
condition is satisfied and the geometry is regular, it is easy to
see that, in general, it  will not increase monotonically with $\ell$ except
outside the last apparent horizon.

To show this, we recall that Eq.(4.10) implies

$$
m^\prime= \pi R^2\left[(\rho+J) \omega_+
+ (\rho- J)\omega_-\right]\ge 0\,.\eqno(7.3)$$
The RHS is clearly positive whenever
$\omega_\pm\ge 0$ (or outside the last apparent horizon)
and $\rho\ge |J|$ (see Hayward in Ref.[17]).

If the initial data possesses an apparent horizon,
even though $\rho$ might be large (so that $M$ might also be large
if a singularity does not intervene), if $\rho$ is packed
behind the apparent horizon $m_\infty$ can be arbitrarily small.
We will examine explicit examples in papers II and III.
Physically, the material energy is screened by a large gravitational
binding energy. Because of this the QLM does not provide a very useful
measure of the material energy. The positivity of the
QLM implies that the magnitude of the gravitational binding energy
can never exceed the material energy in a regular geometry.

If our understanding is consistent, our definition of
the gravitational binding energy as the difference between
$m$ and $M$ had better be negative at infinity at least.
It is surprising that, in the maximal slicing, this
inequality holds everywhere. $M$ therefore provides a
global upper bound on $m$,

$$m\le M\eqno(7.4)$$
for all values of $\ell$.
This is true at a MSC where we can always express the
difference [29]

$$M- m=4\pi\int_0^{\ell}d\ell R^2 \rho
\left[1- R^\prime\right]\,.\eqno(7.5)$$
The right hand side is always positive when $\rho$ is
positive because then $R^\prime\le1$ everywhere. In fact, Bizon, Malec and
\'O Murchadha have shown that
we can do much better than this. It is possible to place an extremely
stringent lower bound on the difference in a MSC[30].
We will return to this point in paper II.

In general, the difference is given by

$$\eqalignno{M-m &=\pi \int_0^\ell d\ell R^2
\rho \left[4- (1+J/\rho) \omega_+
- (1- J/\rho)\omega_-\right]\,&(7.6)\cr
&=\pi \int_0^\ell d\ell R^2
\rho \left[(1+J/\rho) 2 - \omega_+)
+ (1- J/\rho)(2 - \omega_-)\right]\,&(7.7)\cr}
$$
When Eq.(5.1) is satisfied and when $K=0$, so that
$|\omega\pm|\le 2$, the difference is positive.
In general, with $K \ne 0$ it is difficult to see how the positivity of $M
- m$ will hold. This is an open question worth settling.

\vfill\eject

\noindent{\bf 8 THE $(\omega_+,\omega_-)$ PLANE}
\vskip1pc

We can exploit the $(\omega_+,\omega_-)$ plane to
represent solutions to the constraints of
spherically symmetric general relativity. We cast the
constraints in the form (2.11a \& b). To solve these equations
we must supplement them with the
equation which anchors $\omega_+$ and
$\omega_-$ to $R$,

$$R'= (\omega_++\omega_-)/4\,.\eqno(2.11c)$$
On the right hand side of (2.11a \& b) appear three
additional functions; the material sources,
$\rho(\ell)$ and $J(\ell)$ and the
extrinsic curvature scalar, $K_{\cal L}(\ell)$.

Our approach has been to specify $\rho(\ell)$ and $J(\ell)$ on
some compact interval $[0,\ell_0]$ consistent with
the dominant energy condition, (5.1), though we have seen that it is
possible to relax this condition to the weak energy condition
under certain circumstances. It is we who decide
whether they do or don't satisfy the energy condition --- this is
not something we derive.

We could specify $K_{\cal L}$ as some function of $\ell$. However,
intuitively, extrinsic curvature should respond to
to the flow of matter $J$. In particular,
if we were to do this, in the
absence of sources, we would find ourselves
foliating Minkowski space non-trivially.
It is therefore not really appropriate to treat $K_{\cal L}$ the same way
as $\rho$ or $J$. What we do is specify some
foliation gauge appropriate to the topology under consideration, (3.1)
say. This permits us to eliminate $K_{\cal L}$ in favor of
$K_R= (\omega_+-\omega_-)/4R$. The right hand sides of
Eqs.(2.11a \&b) now involve only the functions $R$, $\omega_+$ and
$\omega_-$ and the two functions $\rho(\ell)$ and $J(\ell)$.

The resulting three coupled ordinary first order differential equations
can now be solved subject to boundary conditions on $\omega_+$,
$\omega_-$ and $R$ appropriate to the topology, in our case these
boundary conditions are (2.12a \& b) and (2.8a).

Each solution of these equations will define a trajectory
$\Gamma^*\equiv (R(\ell), \omega_+(\ell),\omega_-(\ell))$
on the space of triplets, $(R, \omega_+,\omega_-)$.
What is remarkable is that no essential information
is lost by limiting ourselves to the projections of these
trajectories onto the $(\omega_+,\omega_-)$ plane,
$\Gamma\equiv (\omega_+(\ell),\omega_-(\ell))$.

Whenever the dominant energy condition holds all regular
asymptotically flat trajectories
are bounded on the $(\omega_+,\omega_-)$ plane by
Eqs.(6.2a) and (6.2b) as well as the positivity of the QLM, Eq.(5.2).
Because these inequalities are
independent of $R$ when cast with respect to these
variables, they can be represented in the projection.
The region $\Sigma$ of the plane in which these inequalities are
simultaneously satisfied is indicated in fig.(8.1).
Regular, asymptotically flat solutions are confined to $\Sigma$. Any
trajectory which strays outside $\Sigma$ is necessarily singular.
Not only can it not reenter $\Sigma$, the trajectory
must run off to infinity on the $(\omega_+,\omega_-)$ plane at some
finite value of $\ell$. The details will be discussed in papers II and III.

If the gauge is maximal, this is a square region with
vertices $(2,2)$, $(2,-2)$, $(-2,-2)$ and $(-2, 2)$. The
inequality Eq.(5.2) is a consequence of the other two inequalities.
In general, however, (5.2) bites out two discs
from the square defined by the other two inequalities.
In this paper, the bounds we derive on
$\omega_+$ and $\omega_-$ characterized by the number $\Omega$
do depend on the trajectory itself unless $K=0$.
derived --- not something we put in by hand.
However, the fact that this bound
changes from trajectory to trajectory when $K\ne 0$
is not entirely satisfactory. We can do better. In paper III we show that
in any of the gauges (3.1), $\Omega$ can in turn be bounded by a
universal numerical constant that depends only on the parameter $\alpha$
appearing in the gauge.

The boundary conditions, (2.12) and
(2.13) imply that each non-singular trajectory must both
begin and end on the point, $(2,2)$.
Thus, physical non-singular initial data can be identified with
bounded closed curves, $\Gamma_0$, in $\Sigma$
each of which contains the point $(2,2)$.
In general, trajectories can intersect themselves any
number of times. The configuration space of spherically symmetric
general relativity can be identified as the space of all such bounded
closed trajectories. We note

\item{1.} Vacuum, flat data corresponds to
the zero trajectory, $\Gamma=(2,2)$ for all $\ell$.

\item{2.} Initial data which do not possess
apparent horizons correspond to trajectories
which lie in the upper right hand quadrant,
$\omega_+,\omega_->0$.

\item{3.} Spatial geometries which do not possess extremal surfaces,
$R^\prime>0$, correspond to trajectories which lie to the right of the
principal (negatively sloped) diagonal, $R^\prime=0$

\item{4.} Moment of time symmetry initial data
correpond to trajectories lying on the
positively sloped diagonal, $K_R=0$. The
extremal surface condition coincides with the apparent
horizon condition.

If spacetime is foliated by an $\alpha$
gauge, in addition, we have the inequality, (5.7),
satisfied by $R^\prime$. If $\alpha\ne 2$,
this reduces the range of the allowed trajectories further,
reducing $\Sigma$ to the hexagonal $\Sigma_\alpha$, illustrated
in fig.(9.1).

\item{5.} In an $\alpha$ gauge, when $J\ge 0$ ($J\le 0$)
everywhere so also is $K_R$. Thus the trajectory lies below
(above) the diagonal, $K_R=0$.

\item{6.} The existence of an
extremal surface does not necessarily imply that of a
future apparent horizon. Why this is so
is clear when $J\ge 0$ in an $\alpha$ - gauge.
Conversely, the existence of a
future apparent horizon does not necessarily imply the
existence of an extremal surface.

A small loop in the neighborhood of the point $(2,2)$
corresponds to almost flat initial data.
The length of such a trajectory will typically be close to zero ---
the flat data value. A larger loop,
crossing $\omega_+=0$ has a length bounded from below by $4$.
It would appear that the length of a trajectory corresponds
somehow to how far the initial data is from
vacuum flat initial data. This criterion does not,
however, require the trajectory to venture far from the point $(2,2)$.
For example, a trajectory might wiggle about so much that
it possess an arbitrarily large arc length
even though the distance of maximum excursion from $(2,2)$
is never large.\footnote * {We note that the points
extremizing the length of the excursion from
flat space, $(\omega_+-2)^2 +(\omega_--2)^2$, are the natural
analogs of the points at which $R^{\prime\prime}=0$
at a moment of time symmetry.}
In paper II, we find better criteria to characterize
the distance of the initial data from flat space by
defining a norm on the space of initial data which
consigns singular initial data to infinity --- something the
naive idea proposed here fails to do.

\vskip2pc

\noindent{\bf 9 CONCLUSIONS}
\vskip1pc

This article has been devoted to an examination of the constraints
in spherically symmetric general relativity with the
goal of identifying the physical degrees of freedom.
With this goal in mind, we found that it was useful
to exploit not only the traditional canonical description of the
phase space provided by the metric variables, $(g_{ab}, K_{ab})$,
but also the optical scalar variables, $(\omega_+,\omega_-)$.
An intriguing feature of the optical scalars is
the possibility of casting the constraints in the linear form
(2.11a and b). Working with the appropriate set of variables, we could
focus in on different properties of the phase space; $g_{ab}$ to
describe spatial metric properties;  $K_{ab}$ to describe
the foliation; $\omega_+$ and $\omega_-$ to describe the light
cone. Solutions are represented as trajectories on the $(\omega_+,\omega_-)$
(or $(R',RK_R)$) plane as described in Sect.8.

Our approach distinguishes between properties which are
spacetime diffeomorphism invariant --- independent of the foliation, and
those which are not.

When we do fix the foliation, we take particular care to
introduce gauges which are global and which do not break down
at either minimal surfaces or at apparent horizons.
We introduced a one-parameter family of linear extrinsic time foliations
of spacetime in Sect.3 which includes both
the polar gauge and the maximal slicing condition.
It turns out that only a subset of these give reasonable asymptotic
falloffs to the initial data and these gauges are those bounded by the null
directions of the superspace metric. These gauges are
the natural asymptotically flat foliations of spacetime.
If the spherically symmetric model is any indication, the null
directions in superspace may provide a guide towards identifying the
natural gauges in less trivial models.

In spherically symmetric general relativity it is easy to identify a
spacetime diffeomorphism invariant
quasi-local mass, which coincides with the Hawking mass, and to write
an expression which relates this quasi-local mass to an
integral over the sources when the constraints are satisfied.
The remarkable feature of the QLM is its positivity when the
geometry is regular and the material sources satisfy certain
very reasonable energy conditions. The positivity of the
QLM appears to suggest the negativity of a physically
realistically defined binding energy.

We found that we could prove the
positivity of the QLM under two different sets of assumptions
on the initial data. One of the proofs is gauge
invariant, the other is not --- relying, in addition, on the
implementation of a valid $\alpha$ - gauge.
However,  whereas the former requires that
matter satisfies the DEC (as one would expect),
the latter does not --- requiring only that matter satisfy
the WEC. This is not to say that the conjunction of the WEC with an
$\alpha$ - gauge is equivalent to the DEC.

Underpinning the positivity in either case,
are various bounds on the canonical variables.
When the former (latter) set of conditions
mentioned in the preceeding paragraph
are satisfied, $\omega_+\omega_-\le 4$
($R^{'2}\leq 1$) everywhere in any regular solution to the constraints.
This bound on $R'$ has a simple geometrical
interpretation in that it demonstrates
that each of the spacelike hypersurfaces
we obtain as solutions to the constraints can be embedded in flat $R^4$.
In fact, given the spherical symmetry, they can be described simply by
curves in $R^2$. Of course, this visualization ignores the nontrivial
extrinsic curvature.

There is no analogous bound on $K_R$ if the DEC is not satisfied.
One might be forgiven for overlook even attempting to
search for such a bound because typically one
does not expect momenta to be bounded.
What is remarkable is that when the DEC is satisfied, $K_R$
is bounded. This is not at all obvious using the metric variables.
The way one proceeds is to establish that both
$\omega_+^2$ and $\omega_-^2$ are bounded
when the DEC is satisfied. It is then a
simple corollary that $K_R$ is also bounded.
Clearly the spacetime light cone structure
on the initial hypersurface, though heavily disguised in the
metric description of the initial data, is encoded in the
constraints. While the guiding principle behind the discovery of the
bounds on $\omega_+$ and $\omega_-$ may be
the positivity of the QLM, these bounds are of a fundamentally different
nature to that on the product,
$\omega_+\omega_-$, which features in the proof of the positivity of the
QLM.
We note that the form of these bounds
is also very different from that of the bounds
separating geometries with apparent horizons from ones which do not.
The existence of bounds on the canonical variables
(or their gradients) appears to be
a very fundamental feature of the spherically symmetric theory,
undermining one's confidence in any naive quantization
of this model which does not take them into account.
It is likely that analogous bounds will exist in the full theory.
Their discovery is a challenge for the future.

What we have learnt about the configuration space of spherically symmetric
general relativity will be built upon in future papers.
In particular, we will show that two ingredients, the description of
solutions as trajectories on the $(\omega_+,\omega_-)$ plane and
the quasi-local mass provide extremely useful practical tools, not
just abstract constructions.
We can exploit the QLM to characterize the behavior of the spatial metric
and the extrinsic curvature in the neighborhood of generic singularities.
The only singularities that can occur in a spherically symmetric
geometry do so because $R$ returns to zero. Generically this
will be accompanied by a divergence of both $R^\prime$
(towards minus infinity) and of $K_R$. No
singular geometries terminate in the region, $\Sigma$,
indicated in fig.(8.1). Where they terminate will depend in an
essential way on the value of the QLM in the neighborhood of the
singularity.

A very brief outline of subsequent papers is:

In II, we will examine the solution of the Hamiltonian constraint
at a moment of time symmetry. In this simplified
setting we can gain useful clues as to how best to
characterize the configuration space of the theory. In III, we extend this
analysis to $J\ne 0$.

\vfill
\eject

\centerline {\bf Acknowledgements}
\vskip1pc
This work was partially supported by Forbairt
grant SC/94/225. We wish to thank Edward Malec, many of the ideas discussed
here originated in discussions with him. We also would like to thank an
anonymous referee for many helpful suggestions.

\vskip2pc
\noindent{\bf Appendix A: Optical Scalars, Apparent Horizons and
Extremal Surfaces}
\vskip1pc
Let us consider a closed two-dimensional spacelike surface, $S$. Each
point on  $S$ possesses two
mutually orthogonal spacetime unit vectors which are normal to $S$.
One of these, $t^\mu$ say, may be taken to be timelike and future
directed. The other vector, $n^\mu$ say, is then spacelike and we
choose it to be pointing outwards.
This choice is clearly well defined up to a Lorentz boost in the
normal tangent space. Alternatively, we can always choose two
null normal vectors, one outward directed and
future pointing, $k^\mu_+$ say,
the other also outward
directed but pointing into the past, $k^\mu_-$ say.

We can represent these null vectors
$$k^\mu_{\pm}=\pm t^\mu+n^\mu\,.\eqno(a.1)$$
$k^\mu_-$ is obtained from $k_+$ by reversing the sign of $t^\mu$.
The divergence of $k^\mu_\pm$ on $S$ is now given by
$$\Theta_\pm=(g^{\mu\nu}+t^\mu t^\nu-n^\mu n^\nu)
\nabla_\mu(\pm t_\nu+n_\nu)\,.\eqno(a.2)$$
Let us suppose that $S$ is embedded in some spacelike
hypersurface ${\cal S}$ with normal vector $t^\mu$. Now the
normal vector to $S$ in ${\cal S}$ is also clearly normal to $t^\mu$
in spacetime. With respect to Gaussian normal coordinates for
spacetime adapted to ${\cal S}$, $t^\mu=(1,0,0,0)$ and $n^0=0=n_0$.
Thus we can express
$$\eqalign{\Theta_\pm =&(g^{ab}-n^a n^b)
\nabla_a(\pm t_b+n_b)\cr
=&\nabla\cdot n \pm (g^{ab}-n^a n^b)K_{ab}\,,\cr}\eqno(a.3)$$
where $g_{ab}$ and $K_{ab}= \nabla_a t_b$ are respectively the
spatial metric and the extrinsic curvature of $\Sigma$.
The second term is the trace of the projection of $K_{ab}$
orthogonal to $S$.
In particular, we observe that $\Theta_+$ and $\Theta_-$ are completely
described by the initial data, $(g_{ab}, K_{ab})$ on the spacelike
hypersurface in which we have embedded $S$. We note that
a change in the prescription of $\Sigma$  will change $\Theta_+$
by a boost factor, $\gamma$ say, and $\Theta_-$ by the factor
$\gamma^{-1}$. Thus neither $\Theta_+$ nor $\Theta_+$ is
a spacetime scalar. However,
their product $\Theta_-\Theta_+$ is. This is the product
which occurs in the quasilocal mass formula
introduced in Sect.5. Clearly, $m$ is also a spacetime scalar.

A future (past) trapped surface is a closed
two-dimensional spacelike surface
on which the divergence of future (past) outward
directed null rays is negative. A future (past)
apparent horizon is the outer boundary of such trapped surfaces.
The appearance of a future (past) apparent horizon signals
(Penrose Theorem[10]) gravitational collapse to form a
black hole (initial conditions which could only have
evolved out of a state which possesses a singularity).
\footnote * {We note that there exists no analog of an
apparent horizon in a Euclidean geometry.}

The condition
$$\nabla\cdot n=0\eqno(a.4)$$
is the condition that the closed two-dimensional spacelike
surface $S$ be an extremal hypersurface of ${\cal S}$.
The apparent horizon coincides with an extremal (minimal)
surface in a MSC, regardless of the material content of the theory.

In the spherically symmetric model, the normal spacelike vector
to the spherically symmetric two-dimensional surface of
circumferential radius $R$ is given by $n^a={1\over {\cal L}}(1,0,0)$
and
$$\nabla_a n^a={2\over R}R^\prime\,.$$
In addition,
$$(g^{ab}-n^a n^b)K_{ab}=2 K_R\,.$$
This term therefore not only vanishes in a MSC but also
when $K_R=0$. In general, however, it will not.
We can write

$$\Theta_\pm=  {2 \over R} \Big(R^\prime \pm  R K_R\Big)\,.\eqno(a.5)$$
Recall that with respect to the proper timelike normal,
$RK_R =\dot R$. We can therefore alternatively
write

$$\Theta_\pm=  {2 \over R} k^\mu_{\pm} \nabla_\mu R\,.$$
In flat space foliated by
flat spacelike hypersurfaces, $R=\ell$ independent of $t$,
and $K_{ab}=0$. Thus
$$\Theta_\pm R = 2\,,$$
for all $R$.

A minimal surface in the spatial geometry
does not necessarily correspond to any physically significant
locus of points on the spatial geometry.
However, it does imply that the
geometry possesses either a future or a past apparent horizon.
Even if an apparent horizon is not
present on the initial spacelike surface,
as the system evolves an apparent horizon
might form. One of the nice things about the
identification of the radial coordinate with the
proper radius is that it is insensitive to the formation of
minimal surfaces or trapped surfaces.
This should be contrasted with Schwarzschild coordinates, which
even if globally valid on the initial surface,
will not necessarily remain so.

\vskip2pc

\noindent{\bf Appendix B: Spacetime Approach to
Eqs.(2.11a \& b) }

\vskip1pc

Another route to the derivation of Eqs.(2.11a) and (2.11b)
which makes them, perhaps, more obvious is the following spacetime
approach suggested to the authors by the referee. We note that
the constraint equations (2.1a) and (2.1b) are equivalent to the
projected Einstein equations ($G^{\mu\nu}$ is the Einstein tensor)

$$G^{\mu}{}_\nu t^\nu = 8\pi T^\mu{}_\nu t^\nu\,,\eqno(b.1)$$
where $t^\mu$ denotes the future directed unit normal to the
hypersurface, ${\cal S}$. We exploit the `radial' Einstein equation [9]

$$h^\alpha{}_\mu h^\beta{}_\nu\nabla_\alpha\nabla_\beta R =
{m\over R^2} h_{\mu\nu} -4\pi R T^{\alpha\beta}\epsilon_{\alpha\mu}
\epsilon_{\beta\nu}\,,\eqno(b.2)$$
where $h^{\mu\nu}$ is the $r-t$ part of the spacetime metric, $m$
is the QLM

$$m := {R\over 2}
\left(1-\nabla_\nu R \nabla^\nu R\right)$$
discussed in sect.4, and $\epsilon_{\mu\nu}$ is a
2-form associated with the surfaces orthogonal to the
orbits of the rotation group. If $n^\mu$ be the radial tangent to
${\cal S}$, then $\epsilon_{\mu\nu} = 2 t_{[\mu} n_{\nu]}$.
By contracting (b.2) with $n^\beta$ we obtain a set of equations
equivalent to (b.1). The Hamiltonian and momentum
constraints obtain by projecting
the resulting equations onto $t^\alpha$ and $n^\alpha$
respectively. If, however, we project onto the
two linear combinations $k_\pm^\alpha $ defined by Eq.(a.1),
we get

$$k_\pm^\alpha n^\beta \nabla_\alpha\nabla_\beta R =
{m\over R^2} \mp  4\pi R T_{\alpha\beta}
k_\pm^\alpha t^\beta\,,\eqno(b.3)$$
a set of equations equivalent to Eqs.(2.11a) and (2.11b).

\vskip2pc
\centerline{\bf Figure Caption}
\vskip1pc
\noindent {\bf fig.(8.1)} The $(\omega_+,\omega_-)$ plane.
Regular asymptotically flat solutions are confined to the region,
$\Sigma$, bounded by the closed union of line and arc segments
$AB$,$BC$,$CD$,$DE$,$EF$, and $FA$. When Eq.(3.1) is satisfied,
this is reduced to the hexagonal region $\Sigma_\alpha$,
bounded by the closed union of line segments,
$A^\prime B^\prime$,$B^\prime C$,$CD^\prime$,
$D^\prime E^\prime$,$E^\prime F$, and $FA^\prime$.
Both a regular trajectory $\Gamma_0$, and a singular one
$\Gamma$ are illustrated.

\vfill
\eject
\centerline {\bf REFERENCES}
\vskip1pc
\item{1.} R. Arnowitt, S.Deser and C. Misner in
{\it Gravitation: An Introduction to Current Research},
ed. by L. Witten (Wiley, New York, 1962)
See also
C. Misner, K. Thorne, J. Wheeler {\it Gravitation}
(Freeman, San Fransisco, 1972) Chapter 21
\vskip1pc
\item{2.} S. Hojman, K. Kucha\v r and C. Teitelboim,
{\it Ann. Phys.(N.Y.)} {\bf 98}, 88 (1976);
V. Mukhanov and A. Wipf
{\it On the Symmetries of Hamiltonian Systems}
(Preprint ETH-TH/94-04, 1994)
\vskip1pc
\item{3.}
Y. Choquet-Bruhat and J. York in {\it General Relativity and
Gravitation} ed. by A. Held (Plenum Press, New York, 1980)
\vskip1pc
\item{4.} B. Berger, Chitre, V. Moncrief and Y. Nutku
{\it Phys. Rev} {\bf D 5 } 2467 (1972)
\vskip1pc
\item{5.} W. Unruh {\it Phys Rev} {\bf D 14} 870 (1976)
In this work, a serious error in ref.[5] is corrected.
\vskip1pc
\item{6.} P. Hajecek {\it Phys Rev} {\bf D 26} 3384 (1982)
\vskip1pc
\item{7.} For some recent work treating the vacuum
Schwarzschild spacetime from a Hamiltonian point of
view, see K. Kucha\v r {\it Phys. Rev.} {\bf D50} 3961 (1994)
H. Kastrup and T. Thiemann {\it Nucl. Phys.} {\bf B425} 665 (1994)
\vskip1pc
\item{8.} M. Choptuik, {\it Phys. Rev. Lett.} {\bf 70} 9 (1993)
\vskip1pc
\item{9.} G. A. Burnett {\it Phys Rev} {\bf D 43} 1143 (1991)
\vskip1pc
\item{10.} S.W. Hawking and G. Ellis {\it The Large Scale Structure of
Spactime} (Cambrige Univ. Press, Cambridge, 1973)
\vskip1pc
\item{11.} S.W. Hawking {\it Comm. Math. Phys} {\bf 43} 199 (1975)
\vskip1pc
\item{12.} E. Farhi, A. Guth and J. Guven {\it Nucl. Phys} {\bf B339} 417
(1990)
\vskip1pc
\item{13.} W. Fischler, D. Morgan and J. Polchinski,
{\it Phys. Rev} {\bf D41} 2638 (1990).
\vskip1pc
\item{14.} As a linear combination of intrinsic and extrinsic
geometrical quantities these variables recall
Ashtekar's connection variables.
See, A. Ashtekar {\it New Perspectives in Canonical
Gravity} Bibliopolis, Naples, 1988)
\vskip1pc
\item{15.} E. Malec and N. \'O Murchadha, {\it Phys. Rev.} {\bf 50}
R6033 (1994)
\vskip1pc
\item{16.} B. DeWitt {\it Phys. Rev} {\bf 160} 1113 (1966)
\vskip1pc
\item{17.} The notion of a quasi-local mass in spherically symmetric
general relativity apparently dates back at least as far as R.C. Tolman.
{\it Relativity, Thermodynamics and Cosmology}
(Clarendon Press, Oxford, 1934). See also
C.W. Misner and D.H. Sharp, {\it Phys. Rev.}
{\bf 136B} 571 (1964) and M. A. Podurets, {\it Soviet Astronomy} {\bf
8} 19 (1964); More recently, it was exploited by  P. S. Wesson an J.
Ponce De Leon {\it Astron. Astrophys.} {\bf 206} 7 (1988); W. Fischler
et al. in Ref.[13]; E. Poisson and W. Israel, {\it Phys. Rev.} {\bf
D41} 1796 (1990); G.A. Burnett, {\it Phys. Rev.} {\bf D48} 5688 (1993)
(see also Ref.[9]); S.A. Hayward, gr-qc/9408002 See also J. Brown and
J. York {\it Phys. Rev} {\bf D 47} 1407 (1993) and S.A. Hayward, {\it
Phys. Rev.} {\bf D49} 831 (1994) for further references.
\vskip1pc
\item{18.} K. Kucha\v r {\it Time and Interpretations
of Quantum Gravity}, (Utah Preprint, 1991)
\vskip1pc
\item{19.} E. Malec and N. O' Murchadha, {\it Phys. Rev.} {\bf 49}
6931 (1994)
\vskip1pc
\item{20.} J. Guven and N \'O Murchadha, (gr-qc/9411010, 1994)
subsequently referred to as II.
\vskip1pc
\item{21.} J. Guven and N \'O Murchadha, (Preprint, 1995)
subsequently referred to as III.
\vskip1pc
\item{22.} See, for example, N. \'O Murchadha {\it J. Math Phys} {\bf
27}, 2111(1986)
\vskip1pc
\item{23.} R. Beig and N. \'O Murchadha, (preprint 1995)
\vskip1pc
\item{24.} See, for example, A. Ashtekar in
{\it Lectures in honor of C. Misner} ed. by B.L. Hu and M. Ryan Jr.
(Cambridge University Press, Cambridge, 1993)
\vskip1pc
\item{25.} S. Hawking {\it J. Math Phys} {\bf 9} 598 (1968)
\vskip1pc
\item{26.} R. Schoen and S.T. Yau  {\it Comm Math Phys} {\bf 65} 45
(1979), {\bf 79} 231 (1981); E. Witten {\it Comm Math Phys} {\bf 80}
381 (1981)
\vskip1pc
\item{27.} L. Eisenhart {\it Riemannian Geometry}
(Princeton Univ. Press, Princeton, New Jersey 1925)
\vskip1pc
\item{28.} R. Arnowitt, S.Deser and C. Misner {\it Phys. Rev.}
{\bf 120}, 313 (1960)
\vskip1pc
\item{29.} See, for example, R.M. Wald, {\it General Raltivity}
(Chicago, 1984)
\vskip1pc
\item{30.} P. Bizo\'n, E. Malec and N. \'O Murchadha {\it Class Quantum
Grav} {\bf 7}, 1953 (1990)

\bye